\documentclass[a4paper]{article}
\usepackage{amsmath}
\usepackage{amsfonts}
\usepackage{float}
\usepackage{bbm}
\usepackage{bm}
\usepackage{listings}
\usepackage{graphicx}
\usepackage{color}
\usepackage[font={small}]{caption}

\usepackage[linktocpage,hidelinks]{hyperref}
\usepackage{setspace}

\usepackage{caption}
\usepackage{subcaption}
\usepackage{ulem}
\usepackage{algorithm}
\usepackage[noend]{algpseudocode}

\usepackage{natbib}

\makeatletter
\def\BState{\State\hskip-\ALG@thistlm}
\makeatother

\DeclareMathOperator*{\Ex}{E}
\DeclareMathOperator*{\Var}{Var}

\DeclareMathOperator*{\MAR}{MAR}
\DeclareMathOperator*{\tMAR}{tMAR}

\title{Bayesian mixture autoregressive model with Student's t innovations}

\author{Davide Ravagli \\ \href{davide.ravagli@manchester.ac.uk}{\normalsize davide.ravagli@manchester.ac.uk} 
	\and Georgi N. Boshnakov \\ \href{georgi.boshnakov@manchester.ac.uk}{\normalsize
          georgi.boshnakov@manchester.ac.uk}
      }

\date{%
  \small
  Department of Mathematics, 
  The University of Manchester \\%
  Oxford Road, Manchester M13 9PL, United Kingdom\\[2ex]%
}
                                        
\begin{document}

\maketitle
\begin{abstract}
This paper introduces a fully Bayesian analysis of 
mixture autoregressive models with Student t components.
With the capacity of capturing the behaviour in the 
tails of the distribution, 
the Student t MAR model provides a more flexible modelling
framework than its Gaussian counterpart, leading to
fitted models with fewer parameters and of easier 
interpretation. The degrees of freedom are also treated
as random variables, and hence are included in the 
estimation process.
\end{abstract}
\section{Introduction}
Mixture autoregressive (MAR) models \citep{WongLi2000} were introduced as a flexible
tool to model time series data which presents asymmetry, multimodality
and heteroskedasticity. For this reason, MAR models have proven valid
to deal with financial returns, which often present one or more 
of such features. 

In their paper, Wong and Li describe a MAR model with Gaussian innovations,
in which the condtitional distribution of each component in the mixture 
is assumed to be Normal, using the EM-Algorithm \cite{dempster1977maximum} 
for paramter estimation. 
Since this, several examples of Bayesian estimation for 
MAR models with Gaussian innovations have been presented, see for
instance \citet{sampietro2006}.

\cite{wong2009student} introduced the mixture autoregressive model with
Student t innovations, in which the mixture components are now assumed,
conditionally on the past history of the process, to follow a Student's
t distribution. The reason behind this different hypothesis for the 
components is that the Student t distribution, having heavier tails
than the Normal distribution, would be more suitable to model
financial returns. In addition, it was argued by \cite{wong2009student}
that, because the tails of the distribution can be adjusted, a higher
level of flexibility is achieved compared to the Gaussian MAR model.

We present here a fully Bayesian approach to estimating paramters
of a mixture autoregressive model with Student t innovations.
In particular, conditional to the past history of the process,
each mixture component is assumed to follow a standardised Student
t distribution. In this way, component variances do not depend 
on the degrees of freedom, so that they can be estimated directly.
The proposed method is able to identify the best model to
fit a time series, as well as estimate parameter posterior 
distributions. 
 
The degrees of freedom of each mixture component are treated as 
a parameter in the model. In the Bayesian framework, \citet{gewekeLinear}
proposes a suitable prior distribution for such parameters in the
case of a linear regression model with Student t errors. 
However, it appeared that results are highly affected by the
choice of prior distribution, and therefore one must be careful
incorporating their prior belief or knowledge about the data.
\cite{geweke1994} also used a similar approach to time series
data with the assumption of Student t innovations.

In general, it is conventient for the Student t distribution 
to constrain the degrees of freedom parameters to be larger than $2$,
to ensure existence of both first and second moments. 
\citet{gewekeLinear} and \citet{geweke1994}, as well as different
apporches to the problem such as \citet{fonseca2008}, do not seem
to take this into account in their analysis. For this reason,
we propose a prior distribution for the degrees of freedom 
that ensures existence of the first and second moment.

The paper proceeds as follows: Section 2 reviews the mixture
autoregressive model with Student t innovations, its properties,
the missing data specification
and the first and second order stationarity condition.
Section 3 presents a fully Bayesian analysis of the 
MAR model with Student t innovations, including model
selection and estimation of parameter posterior distributions.
Section 4 shows a simulation study to assess the accuracy of the
proposed method, and finally Section 5 presents a real data
analysis.

\section{The mixture vector autoregressive model with Student t innovations}

A process $\lbrace y_t \rbrace $ is said to follow a mixture autoregressive (MAR)
 process with Student t innovations \citep{wong2009student} 
 if its conditional CDF can be written as:

\begin{equation}
F\left(y_t \mid \mathcal{F}_{t-1}\right) = \sum_{k=1}^{g} \pi_k 
F_{\nu_k}\left(\dfrac{y_t - \phi_{k0} - 
	\displaystyle \sum_{i=1}^{p_k} \phi_{ki}y_{t-i}}{
 \sigma_k}\right)
\label{eq:CDF}
\end{equation}
where:
\begin{itemize}
\item $\mathcal{F}_{t}$ is the sigma-field generated by the process up to, and 
including (t-1).

\item $g$ is the number of mixture components.

\item $\pi_k > 0, k=1,\ldots,g$ are the mixing weights, specifying a 
discrete probability distribution in $[1, g]$ such that $\sum_{k=1}^{g}\pi_k=1$.

\item $F_{\nu_k}(\cdot)$, $k=1,\ldots,g$ denotes the conditional CDF of a standardised
Student t distribution for component $k$ of the mixture, with corresponding 
degrees of freedom $\nu_k$. Formally, we denote a standardised t distribution
with mean $\mu$, variance $\sigma^2$ and degrees of freedom $\nu$ as
$\mathcal{S}(\mu, \sigma^2, \nu)$.

\item  $\phi_k = \left(\phi_{k0}, \phi_{k1}, \phi_{kp_k} \right)$ is the
vector of autoregressive parameters for the $k^{th}$ mixture component,
with $\phi_{k0}$ being shift parameter. $p_k$ is the autoregressive order,
and we $p=\max(p_k)$ to be the largest autoregressive order in the model.
A useful convention is to set $\phi_{kj}=0$ for $p_k < j \leq p$.

\item $\sigma_k$, $k=1\ldots,g$ is the scale parameter, and we define 
$\tau_k = 1/\sigma_k^2$, the corresponding "precision" parameter.

\item If the process starts at $t=1$, then \eqref{eq:CDF} holds for $t>p$.
\end{itemize}

For the analysis, we exploit the so called \textit{integral representation}
of the Student t distribution. If a random variable $X$ follows a Student t
distribution with mean $\mu$ and variance $\sigma^2$, 
and degrees of freedom $\nu$, then the marginal pdf of $X$ can be written
as:
\begin{equation}
f_X(x) = \int_{0}^{\infty} f_{X \mid Z}\left(x \mid z\right) 
f_Z(z) dZ
\label{eq:intdt}
\end{equation} 
where $X \mid Z \sim N\left(\mu, \dfrac{\sigma^2}{z}\right)$ and 
$Z \sim Ga\left(\dfrac{\nu}{2}, \dfrac{\nu}{2} \right)$.	
Notice however that this setup is valid for the non-standardised 
Student t distribution,
for which the variance is equal to $\sigma^2 \dfrac{\nu}{\nu-2}$. 
Therefore, for the standardised Student t it is necessary to adjust the
distribution of $Z$ to a $Ga\left(\dfrac{\nu}{2}, \dfrac{\nu - 2}{2} \right)$.
With this adjustment, the variance becomes equal to $\sigma^2$, so it does
no longer depend on the degrees of freedom. At the same time, the degrees
of freedom play a part in the shape of the distribution, including
the tails.

Given \eqref{eq:intdt} and the subsequent considerations, the pdf of the model
can be written as 
\begin{equation}
\begin{split}
f\left(y_t \mid \mathcal{F}_{t-1}\right) &= \sum_{k=1}^{g} \pi_k 
\sqrt{\dfrac{\tau_k \xi_t}{2 \pi}} \exp \Bigg \lbrace -
\dfrac{\tau_k \xi_t}{2} \left(y_t - \phi_{k0} -
\sum_{i=1}^{p+k} \phi_{ki}y_{t-i}\right) \Bigg \rbrace \\ &\times
\dfrac{\dfrac{\nu_k-2}{2}^{\nu_k/2}}{\Gamma \left(\dfrac{\nu_k}{2}\right)}
\xi_t^{\nu_k/2 - 1} \exp \Bigg \lbrace - \dfrac{\nu_k - 2}{2} \xi_t \Bigg \rbrace
\end{split}
\label{eq:pdf}
\end{equation}
where $\xi_t \sim Ga(\dfrac{\nu_k}{2}, \dfrac{\nu_k - 2}{2})$.

\citet{wong2009student} showed that conditional expectation, 
conditional variance and autocorrelation functions are identical to the
Gaussian MAR model. Respectively:
\begin{equation}
\begin{split}
\Ex \left[y_t \mid \mathcal{F}_{t-1} \right] &= \sum_{k=1}^{g}
\pi_k \mu_{tk} \\
\Var \left(y_t \mid \mathcal{F}_{t-1}\right) &= \sum_{k=1}^{g} \pi_k \sigma_k^2
+ \sum_{k=1}^{g} \pi_k \mu_{tk}^2 - \sum_{k=1}^{g} \left(\pi_k \mu_{tk}\right)^2 \\
\rho_h &= \sum_{k=1}^{g} \pi_k \sum_{i=1}^{p} \phi_{ki} \rho_{\lvert h-i \rvert}
,\qquad h \geq 1 
\end{split}
\end{equation} 
where $\mu_{tk} = \phi_{k0} + \sum_{i=1}^{p_k} \phi_{ki} y_{t-i}$ and 
$\rho_h$ is the autocorrelation at lag $h$.

\subsection{Stability of the MAR model}
\label{sec:stability}
A matrix is stable if and only if all of its eigenvalues have moduli smaller
than one (equivalently, lie inside the unit circle).
Consider the companion matrices
\begin{equation*}
	A_k = \begin{bmatrix}
		&\phi_{k1} &\phi_{k2} &\dots &\phi_{k(p-1)} &\phi_{kp} \\
		&1 &0 &\dots &0 &0 \\
		&0 &1 &\dots &0 &0 \\
		&\vdots &\vdots &\ddots &\vdots &\vdots\\
		&0 &0 &\dots &1 &0 \\
	\end{bmatrix}
	, \quad
	k = 1,\ldots,g
	.
\end{equation*}
We say that the MAR model is stable if and only if the matrix
\begin{equation*}
	A = \displaystyle \sum_{k=1}^{g} \pi_k A_k \otimes A_k
\end{equation*}
is stable ($\otimes$ is the Kronecker product).
If a MAR model is stable, then it can be used as a model for stationary time
series. The stability condition is  sometimes called stationarity condition, as when this condition holds, 
the model is guaranteed to be first and second order
stationary (i.e. weakly stationary). 

If $g = 1$, the MAR model reduces to an AR model and the above condition states
that the model is stable if and only if $A_1 \otimes A_1$ is stable, which is
equivalent to the same requirement for $A_1$.
For $g > 1$, it is still true that if all matrices $A_{1},\dots,A_{g}$,
$k=1,\dots,g$, are stable, then $A$ is also stable. However, the inverse is no
longer true, i.e. $A$ may be stable even if one or more of the
matrices $A_{k}$ are not stable.

What the above means is that the parameters of some of the components of a
MAR model may not correspond to stationary AR models. It is
convenient to refer to such components as ``non-stationary''.

Partial autocorrelations are often used as parameters of autoregressive models
because they transform the stationarity region of the autoregressive parameters
to a hyper-cube with sides $(-1,1)$ \citep{BARNDORFFNIELSEN1973408, sampietro2006}. 
The above discussion shows that the partial
autocorrelations corresponding to the components of a MAR model cannot be used
as parameters if coverage of the entire stationary region of the MAR model is
desired.

\section{Bayesian analysis of the Student t MAR model}

Given a time series $y_1, \ldots, y_n$, the likelihood function for the 
Student t MAR model using \eqref{eq:pdf} is:
\begin{equation}
\begin{split}
L\left(\boldsymbol{\phi}, \boldsymbol{\sigma}, \boldsymbol{\pi} \mid
\boldsymbol{y}, \boldsymbol{\xi}\right) =
\prod_{t=p+1}^{n} \sum_{k=1}^{g} &
\pi_k 
\sqrt{\dfrac{\tau_k \xi_t}{2 \pi}} \exp \Bigg \lbrace -
\dfrac{\tau_k \xi_t}{2} \left(y_t - \phi_{k0} -
\sum_{i=1}^{p+k} \phi_{ki}y_{t-i}\right) \Bigg \rbrace \\ &\times
\dfrac{\dfrac{\nu_k-2}{2}^{\nu_k/2}}{\Gamma \left(\dfrac{\nu_k}{2}\right)}
\xi_t^{\nu_k/2 - 1} \exp \Bigg \lbrace - \dfrac{\nu_k - 2}{2} \xi_t \Bigg \rbrace
\end{split}
\end{equation}

The likelihood function is not very tractable and a standard approach is to recur to a
missing data formulation \citep{dempster1977maximum}.
Let $Z_t = (Z_{t1},\ldots, Z_{tg})$ be a latent allocation random variable, 
where $Z_t$ is a g-dimensional
vector with entry k equal to 1 if $y_t$ was generated from the $k^{th}$ 
component of the
mixture, and 0 otherwise. We assume that the $Z_{t}$s are discrete random variables, independently
drawn from the discrete distribution:
\begin{equation*}
P(Z_{tk} = 1 \mid g, \boldsymbol{\pi}) = \pi_k
\end{equation*}
This setup, widely exploited in the literature of finite mixture models 
\citep[see, for instance][]{diebolt1994estimation} allows to rewrite the likelihood
function in a much more tractable
way as follows: 
\begin{equation}
\begin{split}
L\left(\boldsymbol{\phi}, \boldsymbol{\sigma}, \boldsymbol{\pi} \mid
\boldsymbol{y}, \boldsymbol{\xi}\right) =
\prod_{t=p+1}^{n} \sum_{k=1}^{g} &
\Bigg (\pi_k 
\sqrt{\dfrac{\tau_k \xi_t}{2 \pi}} \exp \Bigg \lbrace -
\dfrac{\tau_k \xi_t}{2} \left(y_t - \phi_{k0} -
\sum_{i=1}^{p+k} \phi_{ki}y_{t-i}\right) \Bigg \rbrace \\ &\times
\dfrac{\dfrac{\nu_k-2}{2}^{\nu_k/2}}{\Gamma \left(\dfrac{\nu_k}{2}\right)}
\xi_t^{\nu_k/2 - 1} \exp \Bigg \lbrace - \dfrac{\nu_k - 2}{2} \xi_t \Bigg \rbrace
\Bigg ) ^{z_{tk}}
\end{split}
\end{equation}
Notice that, because exactly one $z_{tk}=1$ at each time $t$, the augmented
likelihood is a product, and therefore easier to handle.

In practice, both the $Z_t$s and the $\xi_t$s are not available. We refer them as
latent variables of the model, and we use a Bayesian approach to deal with this.

\subsection{Priors setup and hyperparameters}
\label{sec:priors}
The setup of prior distributions mostly exploits and adapts the existing
literature \citep[for relevant examples see, for instance,][]{diebolt1994estimation,
	gewekeLinear, sampietro2006}. 

In absence of relevant prior information, it is reasonable to assume a priori
that each observation is equally likely to be generated from any of the
mixture components, i.e. $\pi_1 = \dots, \pi_g = 1/g$. This implies a 
discrete uniform distribution for the $Z_t$s, which is a particular case of
the multinomial distribution.
The natural conjugate prior for it is a Dirichlet distribution for 
$\boldsymbol{\pi}$, and therefore we set:

\begin{equation*}
\boldsymbol{\pi} \sim \mathcal{D}\left(w_1, \ldots, w_g \right), \qquad
w_1 = \dots = w_g = 1
\end{equation*}

The prior distribution of each $\xi_t$ directly depends upon the corresponding
$Z_t$, i.e. which of the mixture component $y_t$ was generated from. By model
specification, for
a generic $z_{tk} = 1$, prior distribution on $\xi_t$ is
\begin{equation*}
\xi_t \mid \boldsymbol{z}_t \sim Ga(\dfrac{\nu_k}{2}, \dfrac{\nu_k - 2}{2})
\end{equation*}
The prior distribution on the component means is a Normal distribution with
common hyperparamters $\zeta$ for the mean and $\kappa$ for the precision
\begin{equation*}
\mu_k \sim N\left(\zeta, \kappa^{-1} \right), \qquad k=1,\ldots,g
\end{equation*}

For the precision $\tau_k$, a hierarchical approach is adopted, as 
suggested by \citet{RichardsonGreen1997}. Specifically, we set
\begin{equation*}
\begin{split}
\tau_k &\sim Ga\left(c, \lambda \right), \qquad k=1,\ldots,g \\
\lambda &\sim Ga(a, b)
\end{split}
\end{equation*}

To account for potential multimodality in the distribution, we choose a
multivariate uniform prior distribution for the autoregressive parameters,
limited in the stability region of the model. Hence, for a generic 
$\boldsymbol{\phi}_k$ we have:
\begin{equation*}
p(\boldsymbol{\phi}_k) \propto \mathcal{I}\lbrace \textit{Stable} \rbrace
\end{equation*}
where $\mathcal{I}\lbrace \cdot \rbrace$ is the indicator function assuming 
value 1 if the model is stable and 0 otherwise.

For prior distribution on the degrees of freedom $\nu_k$, $k=1,\ldots,g$, 
\citet{gewekeLinear} suggests an exponential distribution. However, the
posterior distribution could potentially be highly influenced by the choice
of prior, and therefore choosing an exponential prior could result in favour
of low degrees of freedom.
We opt instead for a 
$Ga(\alpha_k, \beta_k)$, $k=1,\ldots,g$ 
prior distributions, which are more flexible, and allow to
better incorporate prior information or belief. 

Two more considerations have to be made: the degrees of freedom parameter 
must assume value larger than $2$ for existence of first and second moments of
the Student t distribution; For degrees of freedom larger than $30$, it is 
reasonable to use a Normal approximation. Therefore, we opted for truncating
the prior distribution so that only values in the interval $[2,30]$ belong
to the parameter space.

\paragraph*{Choice of hyperparameters}

We require specification for the hyperparameters $\zeta$, $\kappa$, $c$, $a$ and $b$.
Although $\lambda$ is also a hyperparameter, it is a random variable, fully 
specified once $a$ and $b$ are chosen.

Following standard setup of mixture models \citep[e.g.]{RichardsonGreen1997},
let $\mathcal{R}_y = \max(y) - min(y)$ be the length variation of the dataset.
Hyperparameters are then set as follows:
\begin{align*}
a&=0.2 & c&=2  &b=\dfrac{100a}{c\mathcal{R}_y^2} = \dfrac{10}{\mathcal{R}_y^2} \\
\zeta &= \min(y) + \dfrac{R_y}{2} &\kappa &= R_y^{-1} &\\
\end{align*}

The choice of $\alpha_k$ and $\beta_k$ for prior distributions
of degrees of freedom parameters are the result of the following reasoning:
\begin{itemize}
\item In general, choosing $\alpha>1$ ensures a peak in the gamma 
distribution, which
could drive the posterior distribution towards such peak.
\item The mode of a gamma distribution is equal to 
$\dfrac{\alpha - 1}{\beta}$, because of the inevitable subjectivity of this prior,
it is reasonable to choose a distribution 
that sees its peak around the point of maximum likelihood. 
Denoting $\hat{\nu}_k^{EM}$ the estimate of degrees of freedom using the 
EM-algorithm approach \citep{wong2009student}, we set a condition that
\begin{equation*}
\dfrac{\alpha_k - 1}{\beta} = \hat{\nu}_k^{EM}
\end{equation*}
\item We may  want to assume a priori that degrees of freedom parameters for
all components have a priori the same variance (at least approximately, given the
truncated nature of the prior). Given a target variance $s^2$, 
this can be done by setting:
\begin{equation*}
\dfrac{\alpha}{\beta^2} = s^2 
\end{equation*}
\end{itemize} 
Thus, each $\alpha_k$ and $\beta_k$ are carefully chosen so that these two 
conditions are satisfied.

\subsection{Simulation of latent variables and posterior distributions}
\label{sec:posteriors}
We here give formulas for simulation of the latent variables in the model,
$\boldsymbol{Z}$ and $\boldsymbol{\xi}$, and posterior distributions 
of model parameters.

Let $\boldsymbol{\phi}(\cdot)$ denote the pdf of the standard Normal distribution,
and introduce the following notation:
\begin{align*}
e_{tk} &= y_t - \phi_{k0} - \displaystyle \sum_{i=1}^{p} \phi_{ki}y_{t-i}, & k&=1,\ldots,g; & t &=(p+1), \ldots, n \\
 n_k & = \displaystyle \sum_{t=p+1}^{n} z_{tk} 
 & \bar{e}_k &= \dfrac{1}{n_k}\sum_{t:z_{tk}=1} e_{tk} \\
 b_k & = 1 - \sum_{i=1}^{p} \phi_{ki} 
 & c_k &= \sum_{t:z_{tk}=1} \xi_t \left(e_{tk} - \bar{e}_k \right) \\
  d_k &= \sum_{t:z_{tk}=1} \xi_t
\end{align*}
We have:
\begin{equation}
\begin{split}
P\left(Z_{tk} = 1 \mid \boldsymbol{\pi}, \boldsymbol{\mu}, \boldsymbol{\tau},
\lambda, \boldsymbol{\nu}, \boldsymbol{y}, \xi_t\right) &=
\dfrac{\dfrac{\pi_k}{\sigma_k} \boldsymbol{\phi}\left(
	\dfrac{e_{tk}}{\sigma_k / \sqrt{\xi_t}}\right)}{\displaystyle
	\sum_{l=1}^{g}
\dfrac{\pi_l}{\sigma_l} \boldsymbol{\phi}\left(
\dfrac{e_{tl}}{\sigma_l / \sqrt{\xi_t}}\right)} \\
\xi_t \mid \boldsymbol{z}_t, \boldsymbol{\pi}, \boldsymbol{\mu}, 
\boldsymbol{\tau},
\lambda, \boldsymbol{y}, \boldsymbol{\nu}  &\sim Ga\left(
\dfrac{\nu_k + 1}{2}, \dfrac{\tau_k}{2} e_{tk}^2 + 
\dfrac{\nu_k - 2}{2}\right)\\
\boldsymbol{\pi} \mid \boldsymbol{\mu}, \boldsymbol{\tau},
	\lambda, \boldsymbol{\nu}, \boldsymbol{y}, 
	\boldsymbol{z},\boldsymbol{\xi} &\sim
	\mathcal{D}\left(1 + n_1, \ldots, 1 + n_g\right) \\
	\mu_k \mid \boldsymbol{\pi}, \boldsymbol{\tau},
	\lambda, \boldsymbol{\nu}, \boldsymbol{y}, 
	\boldsymbol{z},\boldsymbol{\xi} &\sim N\left(
	\dfrac{\tau_k b_k (\bar{e}_k d_k + c_k) + \kappa \zeta}{
	\tau_k b_k^2 d_k + \kappa}, \dfrac{1}{\tau_k b_k^2 d_k + \kappa}\right) \\
\tau_k \mid \boldsymbol{\pi}, \boldsymbol{\mu},
\lambda, \boldsymbol{\nu}, \boldsymbol{y}, \boldsymbol{z}, \xi_t & \sim
Ga\left(\dfrac{n_k}{2} + c, \dfrac{1}{2} \sum_{t:z_{tk}=1}
\xi_t e_{tk}^2 + \lambda \right) \\
\lambda \mid \boldsymbol{\pi}, \boldsymbol{\mu}, \boldsymbol{\tau}, 
\boldsymbol{\nu}, \boldsymbol{y}, \boldsymbol{z}, \xi_t & \sim 
Ga\left(a + cg, b + \sum_{k=1}^{g} \tau_k\right)
\end{split}
\end{equation}

Posterior distributions of $\boldsymbol{\phi}_k$ and $\nu_k$ 
do not have the form of a standard
distribution, therefore we recur to Metropolis-Hastings methods for simulation.

For the autoregressive parameters, $\boldsymbol{\phi}_k$, $k=1,\ldots,g$, 
we recur to random walk metropolis. Let $\boldsymbol{\phi}_k$ be the current 
state of the chain. We simulate a candidate value 
$\boldsymbol{\phi}_k^{*}$ from the proposal distribution 
$MVN\left(\boldsymbol{\phi}_k, \gamma_kI_{p_k}\right)$, where $\gamma_k$ is a
tuning parameter and $I_{pk}$ is the $p_k \times p_k$ identity matrix. 
A move to the candidate value $\boldsymbol{\phi}_k^{*}$ is then accepted
with probability
\begin{equation}
\alpha\left(\boldsymbol{\phi}_k^, \boldsymbol{\phi}_k^{*}\right) =
\min \left( 1, \dfrac{\exp \Bigg \lbrace -\dfrac{\tau_k}{2}
	\displaystyle \sum_{t:z_{tk}=1}
	y_t - \phi_{k0}^{*} - \displaystyle
\sum_{i=1}^{p_k} \phi_{ki}^{*} y_{t-i} \Bigg \rbrace}{\exp 
\Bigg \lbrace
-\dfrac{\tau_k}{2} \displaystyle \sum_{t:z_{tk}=1}
y_t - \phi_{k0} - \displaystyle
\sum_{i=1}^{p_k} \phi_{ki} y_{t-i} 
\Bigg \rbrace} \right)
\end{equation}

The posterior distribution of a generic $\nu_k$ can be written as:
\begin{multline}
p\left(\nu_k \mid \boldsymbol{\pi}, \boldsymbol{\mu}, \boldsymbol{\tau},
\lambda, \boldsymbol{y}, \boldsymbol{z}, \boldsymbol{\xi}\right) \propto
\dfrac{\left(\dfrac{\nu_k -2 }{2}\right)^{n_k\nu_k/2}}{
	\Gamma\left(\dfrac{\nu_k}{2}\right)} \prod_{t:z_{tk}=1}\xi_t^{\nu_k/2 - 1}
\exp \lbrace \dfrac{\nu_k-2}{2} \sum_{t:z_{tk}=1}\xi_t \rbrace  \\
\nu_k \exp \lbrace - \beta \nu_k \rbrace
\end{multline}
which is not a standard distribution. 
We propose an independent
sampler. Independently of the current state of the chain, $\nu_k$, we simulate
a candidate value $\nu_k^{*}$ from its prior distribution. In this way, the 
acceptance probability reduces to the likelihood ratio between  the candidate value
 and the current value, i.e.
\begin{equation}
\alpha\left(\nu_k, \nu_k^{*}\right) = 
\min \left( 1, \dfrac{ \left( \dfrac{\nu_k^{*} -2 }{2}\right)^{n_k\nu_k^{*}/2}}{
	\left( \dfrac{\nu_k -2 }{2}\right)^{n_k\nu_k/2}}
\dfrac{\Gamma\left(\dfrac{\nu_k}{2}\right)}{\Gamma\left(\dfrac{\nu_k^{*}}{2}\right)}
\dfrac{\displaystyle \prod_{t:z_{tk}=1}\xi_t^{\nu_k^{*}/2 - 1}}{
\displaystyle \prod_{t:z_{tk}=1}\xi_t^{\nu_k/2 - 1}}
\dfrac{\nu_k^{*}}{\nu_k}
\exp \big \lbrace - \beta (\nu_k - \nu_k^{*})  \big \rbrace \right)
\end{equation}

\subsection{Choosing autoregressive orders}
\label{sec:RJMCMC}
For this step, we recur to reversible jump MCMC \citep{green1995reversible},
updating the equations of Gaussian mixtures to account for
the new
model specification.
At each iteration, one of the $g$ mixture components, say $k$, 
is chosen at random.
Let $p_k$ be the current autoregressive order of such component. In addition,
set $p_{max}$ as the largest possible autoregressive order.
The proposal is to increase the autoregressive order to $p_k^{*} = p_k+1$ with
probability $b(p_k)$, or decrease it to $p_k^{*} = p_k-1$ with probability
$d(p_k)$. $b(\cdot)$ may be any function defined in $[0,1]$ satisfying
$b(p_{max}) = 0$, and $d(p_k) = 1-b(p_k)$. 

Both scenarios have a $1-1$ mapping between current and candidate model, since 
the only difference between the two is the addition or subtraction of the largest
order autoregressive parameter. Therefore, the Jacobian is always equal
to $1$.

Given a proposed move, we proceed as follows:
\begin{itemize}
\item If the proposed move is to $p_k^{*} = p_k - 1$, the autoregressive 
parameter $\phi_{kp_k}$ is dropped from the model, and the acceptance probability
is the product of the likelihood and the proposal ratio, i.e.
\begin{equation}
\alpha(p_k, p_k^{*}) = \min \Bigg \lbrace 1,
 \dfrac{f\left(\boldsymbol{y} \mid \boldsymbol{\phi}_k^{p_k^{*}}\right)}{
 f\left(\boldsymbol{y} \mid \boldsymbol{\phi}_k^{p_k}\right)} \times
\dfrac{b(p_k^{*})}{d(p_k)} \times
\boldsymbol{\phi}\left(\dfrac{\phi_{pk} - \phi_{pk}}{1/\sqrt{\gamma_k}} \right)
\Bigg \rbrace
\end{equation}

\item If the proposal is to move to $p_k^{*} = p_k + 1$, we simulate the 
additional parameter $\phi_{kp_k^{*}}$ from a $\mathcal{U}(-1.5, 1.5)$ 
distribution. This choice ensures that values close to $0$ are equally as 
likely to be taken into consideration as values far from zero, while trying
to maintain the algorithm as efficient as possible in terms of drawing
values within the stability region of the model.

In this case, the acceptance probability is the ratio between the likelihood
and the proposal, i.e.
\begin{equation}
\alpha(p_k, p_k^{*}) = \min \Bigg \lbrace 1,
\dfrac{f\left(\boldsymbol{y} \mid \boldsymbol{\phi}_k^{p_k^{*}}\right)}{
	f\left(\boldsymbol{y} \mid \boldsymbol{\phi}_k^{p_k}\right)} \times
\dfrac{d(p_k^{*})}{b(p_k)} \times 3 \Bigg \rbrace
\end{equation}
where $3$ is the inverse of the density of any $\phi_{kp_k^{*}}$ under
a $\mathcal{U}(-1.5,1.5)$ proposal distribution.
\end{itemize}

Notice that, in both scenarios, if the candidate model does not satisfy the 
stability condition of Section \ref{sec:stability}, then it is automatically
rejected.

Ultimately, the model which is selected the most number of times over a fixed 
number of iterations is retained to be the best fit for the data (for a certain 
fixed $g$).

\subsection{Choosing the number of mixture components}

The analysis presented so far works under the assumption of correct specification
of the number of mixture components $g$. We now need a way to select a suitable
number of mixture components.

Recall the marginal likelihood identity. The marginal likelihood
function, (i.e. only conditional on the number $g$) is defined as:
\begin{equation}
f(y \mid g) = \sum_{p} \int f(y \mid \boldsymbol{\theta}, p, g)
p(\boldsymbol{\theta}, p \mid g) d\theta
\end{equation}
where $\boldsymbol{\theta}$ is the vector of model parameters. In our case,
$\boldsymbol{\theta} = (\boldsymbol{\phi}, \boldsymbol{\mu}, \boldsymbol{\tau},
\boldsymbol{\pi}, \boldsymbol{\nu})$.

For any values $\boldsymbol{\theta}^{*}$, $p^{*}$, $g$ and observed data $y$, 
the marginal likelihood identity can be decomposed into products of quantities
that can be estimated:
\begin{equation}
f(y \mid g) = \dfrac{f(y \mid \boldsymbol{\theta}^{*}, p^{*}, g)
                     p(\boldsymbol{\theta}^{*} \mid p^{*}, g)
                     p(p^{*} \mid g)}{
                 p(\boldsymbol{\theta}^{*} \mid y, p^{*}, g)
                 p(p^{*} \mid y, g)}            
\label{eq:margloglik}            
\end{equation}
Notice that most quantities in \eqref{eq:margloglik} are ready available. In 
fact, $f(y \mid \boldsymbol{\theta}^{*}, p^{*}, g)$ is the conditional
pdf of the data, which is known under the model specification; 
$p(\boldsymbol{\theta}^{*} \mid p^{*}, g)$ is the set of prior densities
on the model parameters (see Section \ref{sec:priors}); 
$p(p^{*} \mid g)$ is the prior on the maximum autoregressive order, which
is discrete uniform in $[1, p_{max}]$ a priori (see Section \ref{sec:RJMCMC});
$p(p^{*} \mid y, g)$ is the posterior distribution of the selected autoregressive
orders, which we approximate by the proportion of times the RJMCMC 
algorithm in Section \ref{sec:RJMCMC} retains such model; finally, 
$p(\boldsymbol{\theta}^{*} \mid y, p^{*}, g)$ is the set of posterior
densities on the model parameters (see Section \ref{sec:posteriors}), which
needs to be estimated.

To estimate $p(\boldsymbol{\theta}^{*} \mid y, p^{*}, g)$ we recur to the  
the methods by \citet{Chib1995} and \citet{ChibJeliazkov2001}, 
respectively for use of output from Gibbs sampling and Metropolis-Hastings
sampling. 
The method is analogous to that used in \citet{sampietro2006},
taking into account the different model specification, and
the additional model parameters introduced for the degrees of freedom of 
each mixture component.

Notice that $p(\boldsymbol{\theta}^{*} \mid y, p^{*}, g)$ can be further 
decomposed into a product:

\begin{equation}
\begin{split}
p(\boldsymbol{\theta}^{*} \mid y, p^{*}, g) = 
&p(\boldsymbol{\phi}^{*} \mid y, p^{*}, g) \\
&p(\boldsymbol{\nu}^{*} \mid 
   \boldsymbol{\phi}^{*}, y, p^{*}, g) \\
&p(\boldsymbol{\mu}^{*} \mid  
   \boldsymbol{\phi}^{*}, 
   \boldsymbol{\nu}^{*},y, p^{*}, g) \\
&p(\boldsymbol{\tau}^{*} \mid 
   \boldsymbol{\phi}^{*},
   \boldsymbol{\nu}^{*},
   \boldsymbol{\mu}^{*}, y, p^{*}, g) \\
&p(\boldsymbol{\pi}^{*} \mid 
   \boldsymbol{\phi}^{*},
   \boldsymbol{\nu}^{*},
   \boldsymbol{\mu}^{*},
   \boldsymbol{\tau}^{*}, y, p^{*}, g)
\end{split}
\label{eq:margpost}
\end{equation}

Once all quantities have been estimated, they are plugged into \eqref{eq:margloglik}
to estimate the marginal loglikelihood.

To compare models with different $g$, the
algorithm must be run separately for each individual $g_1, g_2,$ and so on. 
In addition, for better efficiency it is 
recommended that models with different number of mixture components are compared
on the basis of high density values of the parameters according to their
distributions in \eqref{eq:margpost}.

\paragraph*{Estimation of $\boldsymbol{p(\phi^{*} \mid y, p^{*}, g)}$\\}

Posterior distributions of autoregressive paramters are estimated by a
 Metropolis-Hastings algorithm. Here we describe how to estimate the 
 probability of interest.
 
 For a generic mixture component $k$, we partition the parameter space into two
subsets, namely $\Psi_{k-1} = (p, \boldsymbol{\phi}_{1}^{*}, \ldots, 
\boldsymbol{\phi}_{k-1}^{*}, g)$ and $\Psi_{k+1} = (\boldsymbol{\phi}_{k+1},
\ldots, \boldsymbol{\phi}_g, \boldsymbol{\nu}, \boldsymbol{\mu}, 
\boldsymbol{\tau}, \boldsymbol{\pi})$, where parameters in $\Psi_{k-1}$ are
fixed.

First, produce a reduced chain of length $N_j$ for the non-fixed parameters,
and fix $\boldsymbol{\phi}_k^{*}$ to be the highest density value. Define now
$\Psi_{k}=(\Psi_{k-1}, \boldsymbol{\phi}_k^{*})$

Run a second reduced chain of length $N_i$ ($N_i$ and $N_j$ may be equal) for
$\Psi_{k+1}$, as well as a sample $\tilde{\boldsymbol{\phi}}_k$ from the proposal
distribution $MVN(\boldsymbol{\phi}_k^{*}, \gamma_kI_{p_k})$.

Finally, let $\alpha(\boldsymbol{\phi}_k^{(j)}, \boldsymbol{\phi}_k^{*})$ and
$\alpha(\boldsymbol{\phi}_k^{*}, \tilde{\boldsymbol{\phi}}_k^{(i)})$ be
the acceptance probabilities of the Metropolis-Hastings algorithm, respectively
for the first and the second chain. The conditional density at 
$\boldsymbol{\phi}_k^{*}$ can then be estimated as
\begin{equation}
p(\boldsymbol{\phi}_k^{*} \mid \Psi_{k-1}, y, p^{*}, g) = 
\dfrac{\dfrac{1}{N_j} \displaystyle \sum_{j=1}^{N_j}
\alpha(\boldsymbol{\phi}_k^{(j)}, \boldsymbol{\phi}_k^{*})
q_{\phi_k}\left(\boldsymbol{\phi}_k^{(j)},  \boldsymbol{\phi}_k^{*}\right)}{
\dfrac{1}{N_i} \displaystyle \sum_{i=1}^{N_i}
\alpha(\boldsymbol{\phi}_k^{*}, \tilde{\boldsymbol{\phi}}_k^{(i)})}
\end{equation}
where $q\left(\boldsymbol{\phi}_k^{(j)},  \boldsymbol{\phi}_k^{*}\right)$
denotes the density of $\boldsymbol{\phi}_k^{(j)}$ under the proposal
$MVN(\boldsymbol{\phi}_k^{*}, \gamma_kI_{p_k})$.

\paragraph*{Estimation of $\boldsymbol{p(\nu^{*} \mid \phi^{*}, y, p^{*}, g)}$ \\}

Degrees of freedom are also estimated via Metropolis-Hastings, therefore we 
proceed in a similar way.

For a generic component $k$ partition the parameter space into $\Omega_{k-1} =
(p, \boldsymbol{\phi}^{*}, \nu_1, \ldots, \nu_{k-1}, g)$ and $
\Omega_{k+1}=(\nu_{k+1}, \ldots, \nu_g, \boldsymbol{\mu}, \boldsymbol{\tau},
\boldsymbol{\pi})$.

Produce a reduced chain of length $N_j$ for the non-fixed parameters and fix
$\nu_k^{*}$ to be the highest density value, and define $\Omega_k =(\Omega_{k-1},
\nu_k^{*})$. 

Run as second chain of length $N_i$ for $\Omega_{k+1}$, as well as second sample
$\tilde{\nu}_k$ from the proposal disitribution. 
Let $\alpha(\nu_k^{(j)}, \nu_k^{*})$ and $\alpha(\nu_k^{*}, \tilde{\nu}_k^{(i)})$
be acceptance probabilities respectively of the first and second chain.
The conditional density at $\nu_k^{*}$ can be estimated as
\begin{equation}
p(\nu_k^{*} \mid \Omega_{k-1}, y, p^{*}, g) = \dfrac{
\dfrac{1}{N_j} \displaystyle \sum_{j=1}^{N_j} 
\alpha(\nu_k^{(j)}, \nu_k^{*}) 
q_{\nu_k}\left(\nu_k^{(j)}, \nu_k^{*}\right)}{
\dfrac{1}{N_i} \displaystyle \sum_{i=1}^{N_i}
\alpha(\nu_k^{*}, \tilde{\nu}_k^{(i)})}
\end{equation}
where $q_{\nu_k}\left(\nu_k^{(j)}, \nu_k^{*}\right)$ denotes the 
density of $\nu_k^{(j)}$ under the prior (proposal) distribution 
$Ga(\alpha, \beta)$.

\paragraph*{Estimation of $\boldsymbol{p(\mu^{*} \mid \phi^{*}, \nu^{*}, y, p^{*}, g)}$\\}

Run a reduced chain of length $N_j$ for the non-fixed parameters. 
Set $\boldsymbol{\mu^{*}} = (\mu_1^{*}, \ldots, \mu_g^{*})$ 
to be the highest density value.
The posterior density of $\boldsymbol{\mu^{*}}$ can be estimated as:
\begin{equation}
p\left( \boldsymbol{\mu^{*}} \mid \boldsymbol{\phi}^{*}, 
\boldsymbol{\nu}^{*}, y, p^{*}, g\right) =
\dfrac{1}{N} \sum_{j=1}^{N_j} \prod_{k=1}^{g} 
p\left(
\mu_k^{*} \mid \boldsymbol{\phi}^{*}, \boldsymbol{\nu}^{*}, \boldsymbol{\tau}^{(i)},
\boldsymbol{\pi}^{(i)}, y, \boldsymbol{z}^{(i)}, p^{*}, g \right)
\end{equation}

\paragraph*{Estimation of $\boldsymbol{p(\tau^{*} \mid \phi^{*}, \nu^{*}, \mu^{*}, y, p^{*}, g)}$\\}

Run a reduced chain of length $N_j$ for the non-fixed parameters. 
Set $\boldsymbol{\tau^{*}} = (\tau_1^{*}, \ldots, \tau_g^{*})$ 
to be the highest density value.
The posterior density of $\boldsymbol{\tau^{*}}$ can be estimated as:
\begin{equation}
p\left(\boldsymbol{\tau}^{*} \mid \boldsymbol{\phi}^{*}, 
\boldsymbol{\nu}^{*}, \boldsymbol{\mu^{*}}, y, p^{*}, g\right) =
\dfrac{1}{N_j} \sum_{j=1}^{N_j} \prod_{k=1}^{g} 
p\left(
\tau_k^{*} \mid \boldsymbol{\phi}^{*}, \boldsymbol{\nu}^{*}, 
\boldsymbol{\mu}^{*},
\boldsymbol{\pi}^{(i)}, y, \boldsymbol{z}^{(i)}, p^{*}, g \right)
\end{equation}

\paragraph*{Estimation of $\boldsymbol{p(\pi^{*} \mid \phi^{*}, \nu^{*}, \mu^{*}, \tau^{*}, y, p^{*}, g)}$\\}

Run a reduced chain of length $N_j$ for the mixing weights, which are now the 
only non-fixed parameters. 
Set $\boldsymbol{\pi^{*}} = (\pi_1^{*}, \ldots, \pi_g^{*})$ 
to be the highest density value.
The posterior density of $\boldsymbol{\pi^{*}}$ can be estimated as:
\begin{equation}
p\left(\boldsymbol{\pi}^{*} \mid \boldsymbol{\nu}^{*}, \boldsymbol{\phi}^{*},
\boldsymbol{\mu^{*}}, \boldsymbol{\tau}^{*}, y, p^{*}, g\right) =
\dfrac{1}{N_j} \sum_{j=1}^{N_j} 
p\left(
\boldsymbol{\pi}^{*} \mid \boldsymbol{\phi}^{*}, \boldsymbol{\nu}^{*}, 
\boldsymbol{\mu}^{*}, \boldsymbol{\tau}^{*},
y, \boldsymbol{z}^{(i)}, p^{*}, g \right)
\end{equation}

\section{Example}

To illustrate performance of our method, we simulated a time series of length
$n=500$ from the process:
\begin{equation*}
y_t = \begin{cases}
-0.5 y_{t-1} + 0.5 y_{t-2} + \varepsilon_{t1} \qquad &\text{with probability } 0.4 \\
~~ 1.1 y_{t-1} + \varepsilon_{t2} \qquad &\text{with probability } 0.4 \\
-0.4 y_{t-1} + \varepsilon_{t3} \qquad &\text{with probability } 0.2 \\
\end{cases}
\end{equation*}

where $\varepsilon_{t1} \sim \mathcal{S}(0, 5^2, 4)$, 
$\varepsilon_{t2} \sim \mathcal{S}(0, 3^2, 14)$ and
$\varepsilon_{t3} \sim \mathcal{S}(0, 1, 10)$. We denote this as 
$\tMAR(3; 2,1,1)$.

\begin{figure}
\centering
\includegraphics[scale=0.4]{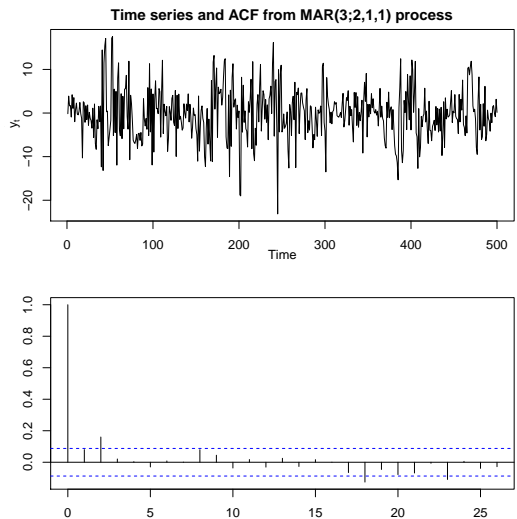}
\caption{Simulated time series from $\tMAR(3;2,1,1)$ process (top) and sample
autocorrelation.}
\label{fig:simul_series}
\end{figure}

The series can be seen in Figure \ref{fig:simul_series}, and it represents what
in practice one should be looking for to assume a MAR model. The series looks
in fact heteroskedastic, amd the plot of the 
sample autocorrelation shows that data are slightly correlated at lag 2. Both
these features may indicate that the underlying generating process is mixture
autoregressive.

 For the analysis, we compared all possible models with 2 and 3 mixture components,
 and maximum autoregressive order equal to 4. 
 
 For what regards the optimal autoregressive orders, the RJMCMC algorithm 
 chooses a $\tMAR(3; 2, 1, 1)$ among all 3-component models, with a preference
 of $0.8054$ (i.e. the model was retained as "best" 3-component model on roughly 
 $81\%$ of the iterations), and a $\tMAR(2; 2, 1)$ among all 2-component models,
 with a preference of $0.8149$.
 When compared with each other in terms of marginal likelihood, 
 the best model was 
 $\tMAR(3; 2, 1, 1)$ with marginal log-likelihood of $-1502.77$ against
 $-1519.166$ for $\tMAR(2; 2, 1)$.
 
\begin{figure}
 	\begin{minipage}{0.6\textwidth}
 \centering
 \includegraphics[height=11cm, width=7.5cm]{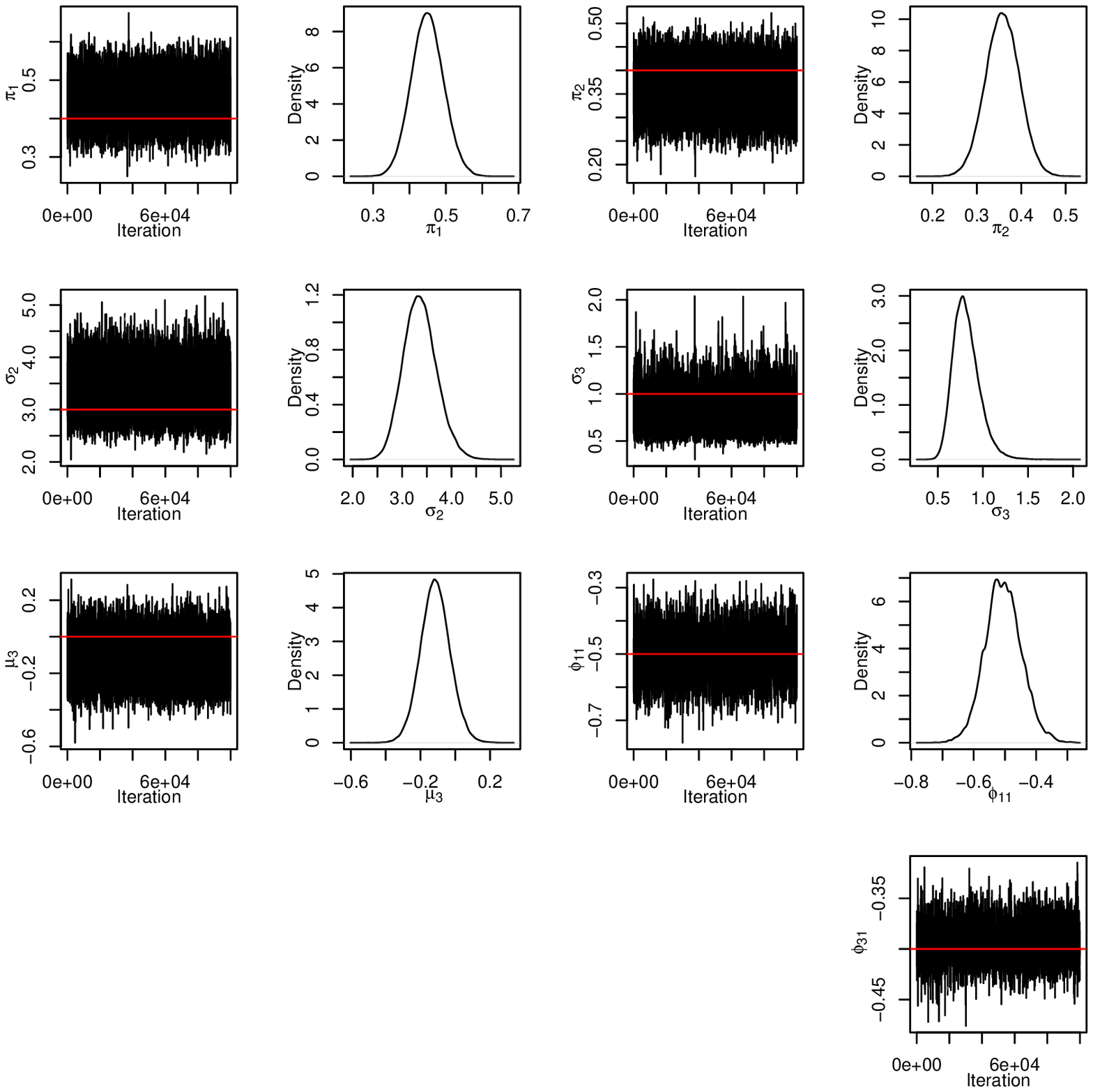}
\end{minipage}
\begin{minipage}{0.6\textwidth}
	\centering
 \includegraphics[height=11cm, width=7.5cm]{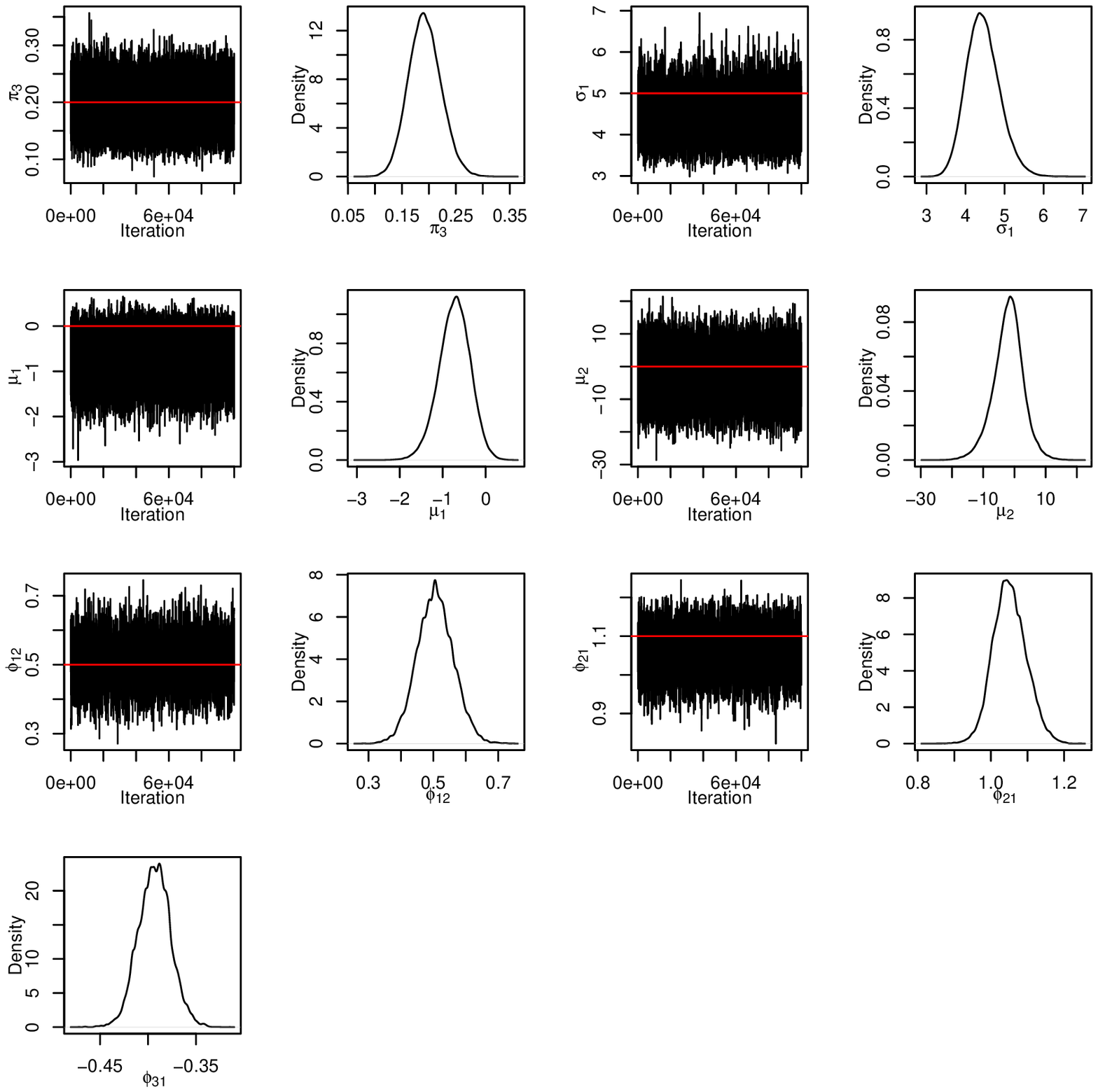}
\end{minipage}
 \caption{Trace and density plots of full conditional posterior distributions 
 	of model parameters under selected $\tMAR(3; 2, 1, 1)$ model. Red lines
 highlight true values.}
\label{fig:simul_densities}
 \end{figure}

\begin{figure}
 \centering
\includegraphics[scale=0.5]{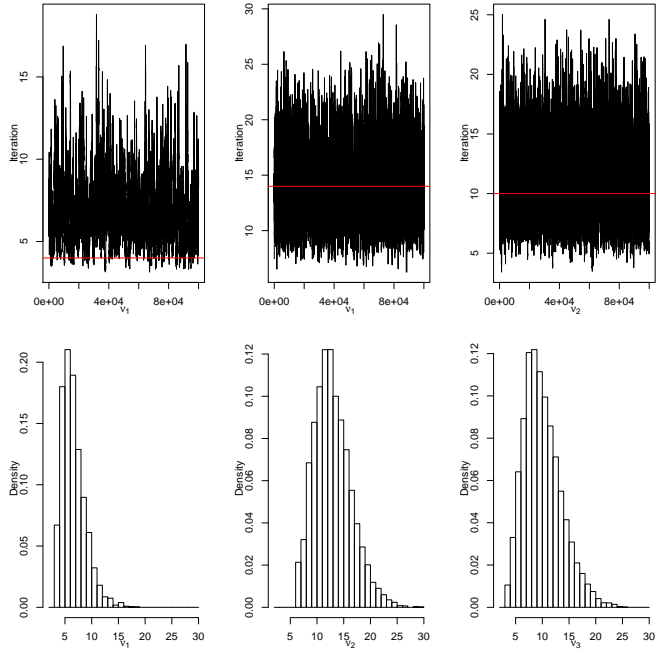}
\caption{Trace plots and histograms of full conditional posterior 
	distributions of degrees
	of freedom parameters under selected $\tMAR(3; 2, 1, 1)$ model, with
unit bin-width. Red lines
highlight true values.}
\label{fig:simul_df_densities}
\end{figure}

 We then simulated a sample of length $100000$ from the posterior distribution
 of the paramters, after allowing $10000$ burn-in iterations. Results
 are displayed in Figure \ref{fig:simul_densities} and 
 Figure \ref{fig:simul_df_densities}. 
 
We can see from Figure \ref{fig:simul_densities} that almost all "true" parameters
are included within the $95 \%$ posterior density region of their respective
distribution. The only exception is found in $\mu_1$, for which such 
region is $[-1.449, -0.0177]$. However, it must be taken into account that
component 1 has the largest variance and the largest autoregressive order,
and is therefore more subject to sampling variability.
For what regards the degrees of freedom parameter, all three components 
have their peak near the true values of the paramaeters: respectively, peaks
are found between $[4,7]$, $[11,13]$ and $[8,11]$ (true values are
$4$, $14$ and $10$).

Overall, we may be satisfied with performance of the algorithm.

\section{The IBM common stock closing prices}

The IBM common stock closing prices \citep{box2015time} is a financial time series
widely explored several times in the literature, including 
\citet{ravagli2020bayesian},
which is our focus for comparison. The series
contains 369 observations from May 17th 1961 to November 2nd 1962.

\begin{figure}[h!]
\centering
\includegraphics[scale=0.4]{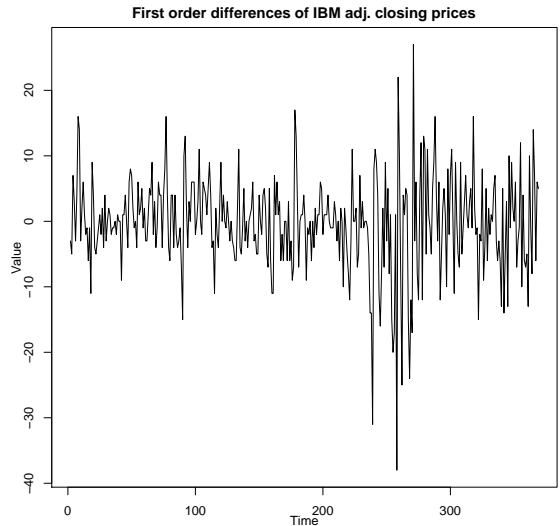}
\caption{Series of first order differences for IBM adjusted closing prices.}
\label{fig:ibm_series}
\end{figure}

We consider the series of first order differences, which can be seen in Figure 
\ref{fig:ibm_series}. The series presents clear signs of heteroskedasticity,
therefore a tMAR model may be a reasonable choice to model the data.

For comparison with previous studies, shift parameters $\phi_{k0}$, $k=1,\ldots,g$
are fixed to 0, hence are not paramters in the model. This taken into account, 
our method chooses a $\tMAR(2; 1, 1)$ as best fit among all tMAR models
with 2 and 3 mixture components and maximum autoregressive order equal to 4.
More specifically, the model was retained about half of the iterations (5067 times
over 10000 iterations) by RJMCMC, meaning it is preferred to models with 2 mixing
 components and larger autoregressive orders.
 Furthermore, the marginal loglikelihood for this model is $-1232.678$, 
 which is larger than that of the competing $\tMAR(3; 2, 1, 1)$, $-1258.073$,
 which was selected as best 3-component model.
 
\begin{figure}
 	\centering
 	\includegraphics[scale=0.5]{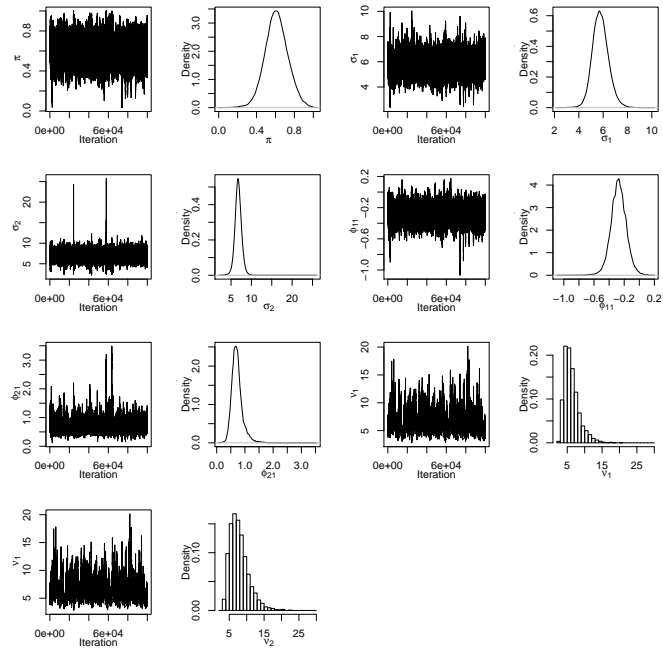}
 	\caption{Trace and density plots of parameter posterior distributions under
 		selected $\tMAR(2; 1, 1)$ model for the IBM data.}
 	\label{fig:ibm_densities}
 \end{figure}

  Once again, we simulated a sample of size 100000 from the posterior distribution
 of the parameters, after 10000 burn-in iterations, which can be seen in Figure
 \ref{fig:ibm_densities}.
 
\citet{ravagli2020bayesian} selected a Gaussian $\MAR(3; 1, 1, 4)$ 
 as best fit for the same dataset, where one of the mixture components was 
 "specialised" to model very few observations with large variability.
 However, the $\tMAR$ model, thanks to its flexibility in the tails of the 
 distribution, only requires 2 components to account for such noise, returning
 a model that is simpler, in that it has fewer parameters, and most importantly
 has a more straightforward interpretation.
 
 \section{Conclusions}
 We have seen a fully Bayesian analysis of mixture autoregressive models with 
 standardised Student t innovations. In a simulation example, it was shown how 
 the method can correctly find the best model to fit a given dataset. In addition,
 we saw that the proposed MCMC for simulation from parameter posterior distributions
 quickly converges to stationarity, and that true values of those parameters 
 are found in high density region.
 
 Secondly, we showed the analysis performed on the IBM common stock closing prices,
 a dataset widely exploited in the literature of heteroskedastic models. In 
 particular, we focused on comparison with the analysis of
 \citet{ravagli2020bayesian} here,
 which used a Gaussian MAR model. Results tell that, thanks to the 
 flexibility of the Student t distribution in its tails, we are able now
 to fit the data with a considerably more parsimonious model, which also has 
 an easier interpretation. 
 
 A limitation of the proposed method is that it inevitably relies on prior 
 information when it comes to degrees of freedom parameters. In practice,
 this means that if one incorporates wrong prior beliefs, the resulting
 posterior distribution will be affected, potentially leading to wrong
 conclusions. One way to make this part of the analysis more objective
 could be adapting Jeffrey's priors for the Student t regression model
 \citep{jeffreyStudent} to the case of mixture autoregression. However,
 this would require derivation of the information matrix.
 On the other hand, the method performs well as long as hyperparameters
 are set "loosely" around the true value of the parameter of interest, 
 so that it may worth using as long as prior information is broadly reliable 
 (i.e. it is sufficient that EM-estimates are available).

\bibliographystyle{kluwer}
\bibliography{stdt_marbayes}

@article{BARNDORFFNIELSEN1973408,
	title = "On the parametrization of autoregressive models by partial autocorrelations",
	journal = "Journal of Multivariate Analysis",
	volume = "3",
	number = "4",
	pages = "408 - 419",
	year = "1973",
	issn = "0047-259X",
	doi = "https://doi.org/10.1016/0047-259X(73)90030-4",
	url = "http://www.sciencedirect.com/science/article/pii/0047259X73900304",
	author = "O Barndorff-Nielsen and G Schou"
}

@misc{ravagli2020bayesian,
    title={Bayesian analysis of mixture autoregressive models covering the complete parameter space},
    author={Davide Ravagli and Georgi N. Boshnakov},
    year={2020},
    eprint={2006.11041},
    archivePrefix={arXiv},
    primaryClass={stat.ME},
    url = {https://arxiv.org/abs/2006.11041}
}

@article{geweke1994, 
	title={Priors for Macroeconomic Time Series and Their Application}, 
	volume={10}, 
	DOI={10.1017/S0266466600008690}, 
	number={3-4}, 
	journal={Econometric Theory}, 
	publisher={Cambridge University Press}, 
	author={Geweke, John}, year={1994}, 
	pages={609–632}
}

@article{WongLi2000,
	author = "Wong, C. S. and  Li, W. K.",
	title = "{On a mixture autoregressive model.}",
	journal = "J. R. Stat. Soc., Ser. B, Stat. Methodol. ",
	volume = 62,
	number = 1,
	pages = "95-115",
	year = 2000,
}

@article{wong2009student,
	ISSN = {00063444},
	URL = {http://www.jstor.org/stable/27798861},
	author = {C. S. Wong and W. S. Chan and P. L. Kam},
	journal = {Biometrika},
	number = {3},
	pages = {751--760},
	publisher = {[Oxford University Press, Biometrika Trust]},
	title = {A Student
	t
	-mixture autoregressive model with applications to heavy-tailed financial data},
	volume = {96},
	year = {2009}
}

@article{dempster1977maximum,
	title = {Maximum likelihood from incomplete data via the EM algorithm},
	author = {Dempster, Arthur P and Laird, Nan M and Rubin, Donald B},
	journal = {Journal of the royal statistical society. Series B (methodological)},
	pages = {1--38},
	year = {1977},
	publisher = {JSTOR}
}

@article{sampietro2006,
	title = {Bayesian analysis of mixture of autoregressive components with an application to financial market volatility},
	author = {Sampietro, S.},
	journal = {Applied Stochastic Models in Business and Industry},
	volume = {22},
	number = {3},
	pages = {242},
	year = {2006},
	publisher = {John Wiley and Sons Ltd.}
}

@article{gewekeLinear,
	ISSN = {08837252, 10991255},
	URL = {http://www.jstor.org/stable/2285073},
	author = {J. Geweke},
	journal = {Journal of Applied Econometrics},
	number = {},
	pages = {S19--S40},
	publisher = {Wiley},
	title = {Bayesian Treatment of the Independent Student-t Linear Model},
	volume = {8},
	year = {1993}
}

@article{fonseca2008,
	ISSN = {00063444},
	URL = {http://www.jstor.org/stable/20441467},
	author = {Thaís C. O. Fonseca and Marco A. R. Ferreira and Helio S. Migon},
	journal = {Biometrika},
	number = {2},
	pages = {325--333},
	publisher = {[Oxford University Press, Biometrika Trust]},
	title = {Objective Bayesian Analysis for the Student-T Regression Model},
	volume = {95},
	year = {2008}
}

@article{diebolt1994estimation,
	title = {Estimation of finite mixture distributions through Bayesian sampling},
	author = {Diebolt, Jean and Robert, Christian P},
	journal = {Journal of the Royal Statistical Society. Series B (Methodological)},
	volume  =  {56},
	pages = {363--375},
	year = {1994},
	publisher = {JSTOR}
}

@article{RichardsonGreen1997,
	author = "Richardson, S. and  Green, P. J.",
	title = "{On Bayesian Analysis of Mixtures with an Unknown Number of Components.}",
	journal = "J. R. Stat. Soc., Ser. B, Stat. Methodol. ",
	volume = "59",
	number = "4",
	pages = "731-792",
	year = "1997",
}

@article{green1995reversible,
	title = {Reversible jump Markov chain Monte Carlo computation and Bayesian model determination},
	author = {Green, Peter J},
	journal = {Biometrika},
	volume = {82},
	number = {4},
	pages = {711--732},
	year = {1995},
	publisher = {Oxford University Press}
}

@article{ChibJeliazkov2001,
	author = "Chib, S. and  Jeliazkov, I.",
	title = "{Marginal likelihood from the Metropolis-Hastings output.}",
	journal = "J. A. Stat. Ass.",
	volume = "96",
	number = "453",
	pages = "270-281",
	year = "2001",
}

@article{Chib1995,
	author = "Chib, S.",
	title = "{Marginal likelihood from the Gibbs output.}",
	journal = "J. A. Stat. Ass.",
	volume = "90",
	number = "432",
	pages = "1313-1321",
	year = "1995",
}

@Book{ box2015time,
	author = { Box, George E. P. and Jenkins, Gwilym M. },
	title = { Time series analysis : forecasting and control / George E.P. Box and Gwilym M. Jenkins },
	edition = { Rev. ed. },
	isbn = { 0816211043 },
	publisher = { Holden-Day San Francisco },
	pages = { xxi, 575 p. : },
	year = { 1976 },
	type = { Book },
	language = { English },
	subjects = { Time-series analysis.; Prediction theory.; Transfer functions.; Feedback control systems -- Mathematical models. },
	life-dates = { 1976 -  },
	catalogue-url = { https://nla.gov.au/nla.cat-vn640184 },
}

@article{jeffreyStudent,
	ISSN = {00063444},
	URL = {http://www.jstor.org/stable/20441467},
	author = {Thaís C. O. Fonseca and Marco A. R. Ferreira and Helio S. Migon},
	journal = {Biometrika},
	number = {2},
	pages = {325--333},
	publisher = {[Oxford University Press, Biometrika Trust]},
	title = {Objective Bayesian Analysis for the Student-T Regression Model},
	volume = {95},
	year = {2008}
}
	
\end{document}